\documentclass[11pt, a4paper]{article}

\usepackage[top=2.5cm, bottom=2.5cm, left=2.5cm, right=2.5cm]{geometry} 
\usepackage{setspace}
\usepackage{amsmath, amssymb, amsfonts, amsthm} 
\usepackage{bm}      
\usepackage{physics}  
\usepackage{siunitx}

\usepackage{graphicx} 
\usepackage{float}    
\usepackage[font=small,labelfont=bf]{caption} 
\usepackage{subcaption}

\usepackage[numbers,sort&compress]{natbib} 
\usepackage[colorlinks=true, linkcolor=blue, citecolor=blue, urlcolor=blue]{hyperref}

\usepackage{authblk}

\title{Phase transition revealed by eigen microstate entropy}

\author[1,2,3,9]{Teng Liu}
\author[1,9]{Xuezhi Niu}
\author[1,9]{Mingli Zhang}
\author[1,4]{Gaoke Hu}
\author[1]{Yuhan Chen}
\author[5]{Yongwen Zhang}
\author[6]{Rui Shi}
\author[6]{Jingyuan Li}
\author[7,8]{Peng Tan}
\author[1,*]{Maoxin Liu}
\author[1,*]{Hui Li}
\author[1,6,*]{Xiaosong Chen}

\affil[1]{School of Systems Science and Institute of Nonequilibrium Systems, Beijing Normal University, Beijing 100875, China}

\affil[2]{Munich Climate Center and Earth System Modelling Group, Department of Aerospace and Geodesy, TUM School of Engineering and Design, Technical University of Munich, Munich, 80333, Germany}

\affil[3]{Potsdam Institute for Climate Impact Research, Potsdam, 14473, Germany}

\affil[4]{College of Physics, Nanjing University of Aeronautics and Astronautics, Nanjing, 211106, China}

\affil[5]{Data Science Research Center, Faculty of Science,
Kunming University of Science and Technology, Kunming 650500, China}

\affil[6]{Institute for Advanced Study in Physics and School of Physics, Zhejiang University, Hangzhou 310058, China}

\affil[7]{State Key Laboratory of Surface Physics and Department of Physics, Fudan University, Shanghai 200433, China}

\affil[8]{Institute for Nanoelectronic Devices and Quantum Computing, Fudan University, Shanghai 200433, China}

\affil[9]{These authors contributed equally to this work.}

\date{}

\begin{document}

\maketitle

{
  \renewcommand{\thefootnote}{\fnsymbol{footnote}}
  \footnotetext[1]{Author to whom any correspondence should be addressed. E-mail: mxliu@bnu.edu.cn, huili@bnu.edu.cn, chenxs@bnu.edu.cn}
}

\begin{abstract}
\noindent
We introduce the eigen microstate entropy ($S_{\text{EM}}$), a novel metric of complexity derived from the probabilities of statistically independent eigen microstates. After establishing its scaling behavior in equilibrium systems and demonstrating its utility in critical phenomena (mean spherical, Ising, and Potts models), we apply $S_{\text{EM}}$ to non-equilibrium complex systems. Our analysis reveals a consistent precursor signal: a significant increase in $S_{\text{EM}}$ precedes major phase transitions. Specifically, we observe this entropy rise before biomolecular condensate formation in liquid-liquid phase separation in living cells and months ahead of El Ni\~no events. These findings position $S_{\text{EM}}$ as a general framework for detecting and interpreting phase transitions in non-equilibrium systems. 
\end{abstract}

\section{Introduction}
As a fundamental concept in physics, entropy has undergone recent advancements, broadening its applications across physical sciences and interdisciplinary fields~\cite{vedral2002role,dong2008exploration,harte2014maximum,ghil2020physics,clare2021physics,li2023cross,dichio2023statistical,nagasawa2025macroscopicity}. These advances underscore the effectiveness of entropy as a measure for characterizing phases of complex systems~\cite{lucarini2009thermodynamic,di2024variance}. However, traditional entropy measures rely on the probability distribution over a system's phase space~\cite{mackey1989dynamic}. While effective for model systems, it becomes intractable for complex systems with multiple components and high-dimensional nonlinear interactions~\cite{tabar2024revealing}.
 
Recent advancements in simulation and observational techniques have yielded extensive phase space data, facilitating the characterization of large-scale behaviors without explicit phase space expressions~\cite{ganguli2012compressed,battle2016broken,thibeault2024low}. These developments have informed various methods related to entropy. Configurational entropy~\cite{berthier2019configurational} and the free energy landscape~\cite{wales2006potential}, for instance, provide insights into how complex systems evolve towards stable states, often by identifying energy minima or calculating energy barriers~\cite{wales2006potential,berthier2019configurational}. Curl flux~\cite{wang2015landscape,xu2020curl} and entropy production~\cite{andrae2010entropy,tome2012entropy,pruessner2025field} utilize microscopic state transition probabilities to extract information on system non-equilibrium effects. The computable information density~\cite{martiniani2020correlation}, as a proxy of entropy, has been applied to capture order and critical behavior in complex systems. While these methods offer valuable perspectives, many are constrained by the requirement of an energy function, specific mathematical formulations, or may not fully account for correlations between sampled configurations. Consequently, we are motivated to develop an entropy measure that addresses these challenges, providing a more versatile approach to understanding complex systems.

In this work, we introduce a data-driven entropy measure, the eigen microstate entropy ($S_{\text{EM}}$). $S_{\text{EM}}$ is based on the probability distribution of eigen microstates, which describes the statistical properties of an ensemble by considering mutually orthogonal eigen microstates within it~\cite{hu2019condensation,zhang2024eigen, li2024exploring,xie2025ecosystem}. This approach offers a possibility to eliminate the ambiguity in entropy definitions caused by statistical correlations among sampled data. We validate $S_{\text{EM}}$ using representative models, including various Ising and Potts models, and demonstrate its advantages by analyzing challenging non-equilibrium systems and experimental scenarios. Specifically, we apply $S_{\text{EM}}$ to address several long-standing problems, including phase transitions in frustrated Ising systems, liquid-liquid phase separation (LLPS), and El Ni\~no prediction.

\section{Eigen microstate entropy}
A microstate of a complex system with $\mathcal{N}$ agent components is represented as a vector $\bm{s}(\tau) = [s_1(\tau),\allowbreak \dots, s_N(\tau)]^{\mathsf{T}}$, where $\tau$ denotes either time tagger or sampling order. By collecting $\mathcal{M}$ samples via experimental observation or simulation, one can construct an $\mathcal{N} \times \mathcal{M}$ ensemble matrix $\bm{\mathrm{A}}$ with entries $A_{i\tau} = {s_i(\tau)}/{\sqrt{C_0}}$, where the normalization factor is given by $C_0=\sum_{\tau=1}^M\sum_{i=1}^Ns_i^2(\tau)$. The so-called eigen microstates $\{ \bm{U}_I\}$ are defined by a singular value decomposition as $\bm{\mathrm{A}} = \sum_{I}\sigma_I\allowbreak \bm{U}_I \otimes \bm{V}_I$~\cite{sun2021eigen,liu2022renormalization}. As its name suggests, $\{ \bm{U}_I\}$ are mutually orthogonal, implying their statistical independence. Any microstate $\bm{s}(\tau) $ can be expressed as a linear superposition of eigen microstates $\{ \bm{U}_I\}$:
$
\bm{s}(\tau)  = \sum_{I=1}^N \sigma_I V_{\tau I}\bm{U}_{I}\sqrt{C_0}.    
$
Given that the dimension $\mathcal{M}$ of the phase space is typically much larger than the dimension $\mathcal{N}$ of the system, the $\mathcal{N}$ mutually independent eigen-microstates offer a dimensionality-reduction approach to reconstructing the statistical ensemble.

With the normalization factor $C_0$ defined above, the sum of the squares of the singular values normalized to $1$, i.e., $\sum_I \sigma_I^2 = 1$. We can thus consider the statistical probability of $\bm{U}_I$ as $p_I = \sigma_I^2$. In this letter, an eigen microstate entropy is introduced as:
\begin{equation}\label{eq:entropy_define}
	S_{\text{EM}} = -\sum_{I=1}^N p_I \ln{p_I}.
\end{equation}
It is expected that the definition of $S_{\text{EM}}$ is not only applicable to model systems but also particularly well-suited for real-world systems, where complete sampling is generally unattainable and the probability of individual samples is therefore difficult to determine. However, $S_{\text{EM}}$ eliminates this ambiguity in probability definition by considering mutually statistically independent eigen microstates. Moreover, the definition of $S_{\text{EM}}$ does not require prior knowledge of the underlying system dynamics, enabling broad application to various complex systems, including those with non-equilibrium characteristics.

\subsection{Finite size scaling of $S_{\text{EM}}$}

We first establish the scaling relation of $S_{\text{EM}}$ to study critical phenomena in theoretical models. Drawing an analogy to the decomposition of the free energy density, we decompose $S_{\text{EM}}$ in the critical regime as: 
\begin{equation}\label{eq:sd}
	S_{\text{EM}}(t,L) =  S_{\text{ns}}(L) + S_{\text{s}}(t,L),
\end{equation}
where $S_{\text{ns}}$ and $S_{\text{s}}$ denote the regular (non-singular) and singular parts of  $S_{\text{EM}}$, respectively. Near a non-zero $T_c$, the regular part is independent of $t$ and scales as: 
\begin{equation}\label{lnL}
    S_{\text{ns}}(L) \propto a\ln{L},
\end{equation}
where $a$ is a constant. Noting that approaching the critical point $T_c$, an emergent phase, $\bm{U}_I$, corresponds to a finite value of  $p_I$, and thus there is a finite-size scaling (FSS) form~\cite{hu2019condensation,sun2021eigen,liu2022renormalization}: $p_I = L^{-2\beta/\nu}\tilde{p}_I(tL^{1/\nu})$, where $L$ is the system size, $t\equiv (T-T_c)/T_c$ is the reduced temperature, $\tilde{p}_I$ is a universal scaling function, and $\beta$, $\nu$ are the critical exponents of the order parameter and correlation length, respectively. To further clarify the microscopic origin of the scaling behavior in $S_{\text{EM}}$, we define the log-ratio $\delta_{I,J}=\ln{(p_I/p_J)}$ as an effective energy gap, where $p_I$ relates to an effective energy via Boltzmann-like relation $p_I = e^{-E_I}/Z$, where $Z$ is an effective partition function. The gap follows FSS form
\begin{equation}
    \delta_{I,J} = \ln[f_I(tL^{1/\nu})/f_J(tL^{1/\nu})],
\end{equation}
where $f_I$ and $f_J$ are universal scaling functions. As this gap is constructed from the $\ln(p_I)$ which is a building block of $S_{\text{EM}}$, it suggests the singular contribution to the entropy, which therefore follows the scaling relation: 
\begin{equation}\label{eq:s}
S_{\text{s}}(t,L) =  L^{-2\beta/\nu} \tilde{S}_s(tL^{1/\nu}),
\end{equation}
where $\tilde{S}_s$ should be a universal scaling function. 

We next analyze the scaling behavior of $S_{\text{EM}}$ in equilibrium systems, deriving analytical results within the mean spherical model and conducting numerical studies on various Ising and Potts models. Furthermore, we extend this analysis to non-equilibrium real-world complex systems, including liquid–liquid phase separation and El Ni\~no events.

\section{Eigen microstate entropy in equilibrium systems}
\subsection{Exact solutions in the mean spherical model}

We analytically solve the mean spherical model in dimensions $2 < d < 4$ under periodic boundary conditions and obtain its eigen microstates. The mean spherical model is described by the Hamiltonian $H=-J\sum_{\langle ij\rangle} s_i s_j + \mu\sum_i s_i^2$, where $J>0$ denotes the nearest-neighbor coupling. This model is analogous to the Ising model but subject to the constraint of spherical field  $\sum_i \langle s_i^2\rangle = N$. Historically, it served as the foundation for M. E. Fisher's formulation of finite-size scaling theory~\cite{barber1973critical}. Here we obtain analytical results of the statistical probability of eigen microstates $\bm{U}_{I}$ (see supplementary data \cite{liu2024SM} for detailed derivations):
\begin{equation}
p_I = L^{-d}\bar\beta^{-1}(\tilde\mu+J_{\mathbf{k}_I})^{-1},
\end{equation}
where $\bar{\beta} = (k_BT)^{-1}$, $\tilde{\mu}=2\mu-2J_0d$, $J_0=2J$, $\mathbf{k}_I=(k_I^1,\ldots,k_I^d)$, and $J_{\mathbf{k_I}}=4J\sum_{j=1}^d(1-\cos{k_I^j})$. In the critical region with $|t|\ll1$ and $L\gg1$, for $|\mathbf{k}_I|=\frac{2\pi}{L} |\mathbf{m}_I| \ll 1$ with finite $|\mathbf{m}_I|$, $p_I$ asymptotically follows the FSS form:
\begin{equation}\label{eq:p_exact}
p_I=L^{-\frac{2\beta}{\nu}}(\bar\beta J_0)^{-1}\!\left[F_{\tilde{\mu}}\!\left(\bar\beta_c J_0\, t\, L^{\frac{1}{\nu}}\right) + 4\pi^2 |\mathbf{m}_I|^2\right]^{-1},
\end{equation}
and the corresponding effective energy gap, 
\begin{equation}\label{eq:msm-energy-fss}
\delta_{I,J} = \ln\!\left[\frac{F_{\tilde{\mu}}\!\left(\bar\beta_c J_0\, t\, L^{1/\nu}\right) + 4\pi^2 |\mathbf{m}_J|^2}
{F_{\tilde{\mu}}\!\left(\bar\beta_c J_0\, t\, L^{1/\nu}\right) + 4\pi^2 |\mathbf{m}_I|^2}\right],
\end{equation}
where $F_{\tilde{\mu}} = J_0^{-1}\tilde\mu L^2$ is a universal scaling function. These modes are responsible for the singular part of $S_{\text{EM}}$. For finite $\mathbf{k}_I$, $p_I = L^{-d}\bar\beta^{-1} J_{\mathbf{k}_I}^{-1}\left[1+\mathcal{O}(L^{-2})\right]$, giving $
\delta_{I,J}=\mathcal{O}(L^{-1})$, which contributes to the non-singular part of $S_{\text{EM}}$.

\begin{figure}[!h]
\includegraphics[width=0.8\textwidth]{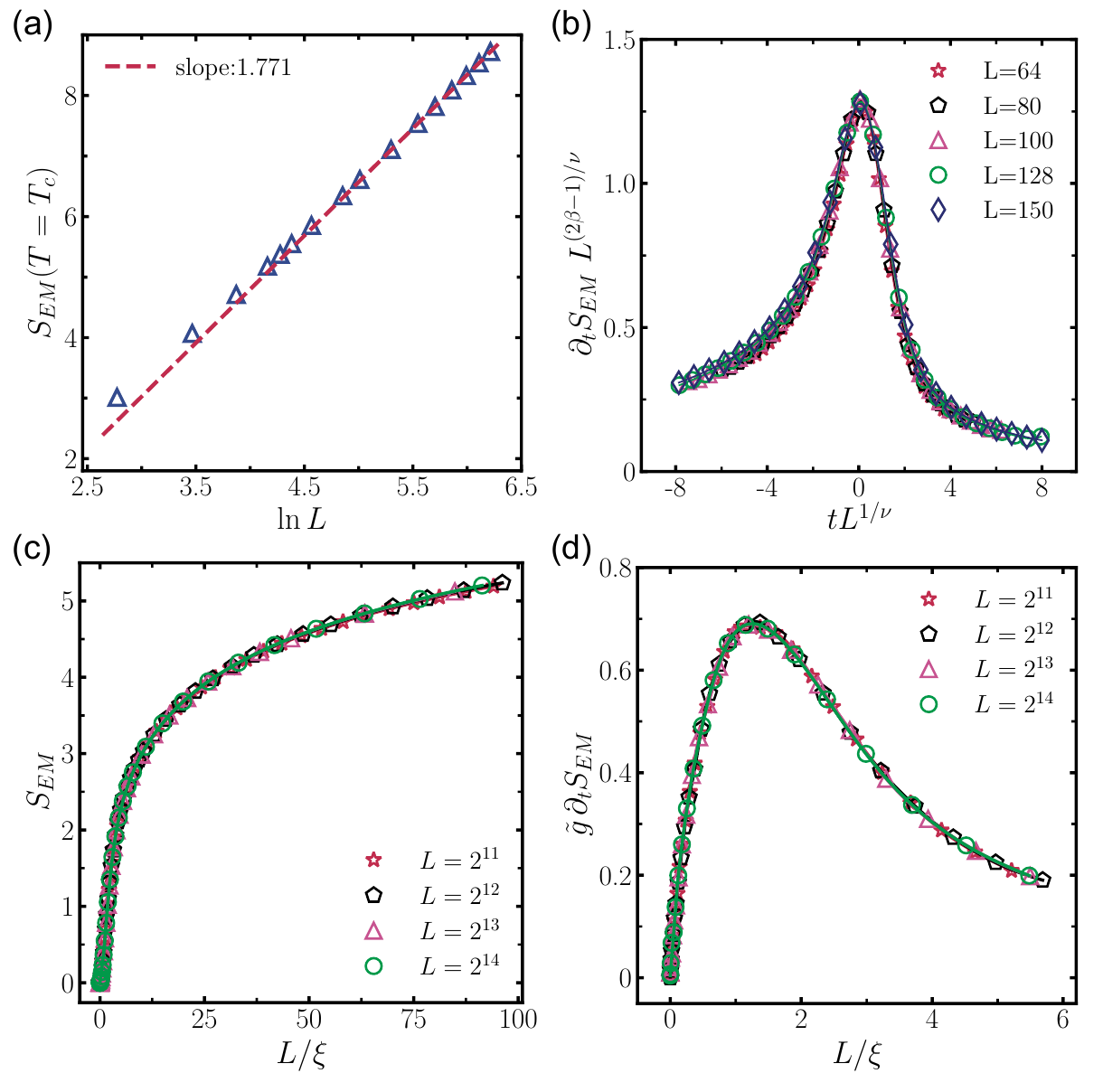}
\centering
\caption{\label{fig:fig1} Critical behaviors of $S_{\text{EM}}$ and its temperature derivative in Ising models. (a) Scaling relation $S_{\text{EM}} \approx S_{\text{ns}}(L) \propto 1.771\ln{L}$ at $T_c$ in the 2D Ising model. Red dashed line represents a fit to the last six data points. (b) Data collapse of $\partial_t{S}_{\text{EM}} L^{(2\beta-1)/\nu}$ versus $tL^{1/\nu}$ in the 2D Ising model with critical exponents $\nu=1$, $\beta=1/8$. (c) Hidden scaling behavior of ${S}_{\text{EM}}$ versus $L / \xi$ in the 1D Ising model. (d) Data collapse of $\tilde{g} \partial_t{S}_{\text{EM}}$ versus $L / \xi$ in the 1D Ising model. Solid lines are guides for the eye; error bars are smaller than symbol sizes.}
\end{figure}

\subsection{Standard Ising model}
Taking the two-dimensional (2D) Ising model as an example, the scaling relation in Eq.\eqref{lnL} is confirmed in figure~\ref{fig:fig1}(a) with a fitted value of $a\simeq1.771(3)$ as a universal amplitude. To eliminate the non-singular part of $S_{\text{EM}}$, we introduce a derivative, $\partial_t{S}_{\text{EM}} = \partial S_{\text{EM}} / {\partial t}$, which serves as a response function. An FSS form is then obtained as 
\begin{equation}\label{eq:ds}
    \partial_t{S}_{\text{EM}}  = L^{(1-2\beta)/\nu}\tilde{K}(tL^{1/\nu}),
\end{equation} where $\tilde{K}$ is a universal scaling function. In the figure~\ref{fig:fig1}(b), the collapse of data from different system sizes into a single curve confirms the scaling relation in Eq.\eqref{eq:ds}. To better fit the scaling relation in Eq.\eqref{lnL} and Eq.\eqref{eq:ds}, critical values $T_c = 2/\ln{(1+\sqrt{2})}$, $\nu =1$ and $\beta=1/8$ can be obtained. It demonstrates the capability of $S_{\text{EM}}$ in studying critical phenomena. 

More interestingly, the above scaling analysis can be applied to the 1D Ising model, in which it has long been acknowledged that there is no phase transition at a non-zero temperature. Approaching zero temperature, the non-singular part $S_{\text{ns}}$ diminishes as $S_{\text{EM}}\rightarrow 0$. Understanding the zero temperature as the critical point of the 1D Ising model, the  behavior of $S_{\text{EM}}$ in the low temperature limit is dominated by the scaling relation in Eq.~\eqref{eq:s}. The correlation length ($\xi = \left (\ln{ [\coth (1/T)]} \right )^{-1}$) diverges as $T \rightarrow 0$~\cite{baxter1982exactly,hu2019condensation}. For the convenience of numerical verification, we alternatively write the scaling relation in Eq.~\eqref{eq:s} as
\begin{equation}\label{eq:s-scaling-1d}
    S_{\text{EM}} \simeq S_{\text{s}} = \tilde{S}_s(L/\xi).
\end{equation} 
This scaling relation is confirmed by the data collapse of  $S_{\text{EM}}$ as a function of  $L/\xi$ across various system sizes (figure~\ref{fig:fig1}(c)). Reminding that the Gibbs entropy in the 1D Ising model can be exactly expressed as 
$
S_{\text{Gibbs}}= N\ln2 + N\ln(\cosh K) + \ln(1 + \tanh^N K) - (NK\tanh K(1 + \tanh^{N-2} K))/(1 + \tanh^N K)   
$
, where $K = 1/T$, which exhibits no explicit scaling features (shown in the supplementary data~\cite{liu2024SM}). In comparison, the Eq.~\eqref{eq:s-scaling-1d} reveals a hidden scaling of $S_{\text{EM}}$. Moreover, an additional hidden scaling is revealed in $\partial_t{S}_{\text{EM}}$, which follows the form $\partial_t{S}_{\text{EM}} = M(L/\xi) / \tilde{g} $, where $\tilde{g} =  \sinh (K)\cosh (K) / L K^2$. This behavior is supported by the numerical calculation (see figure~\ref{fig:fig1}(d)).

\begin{figure}[!h]
\includegraphics[width=0.8\textwidth]{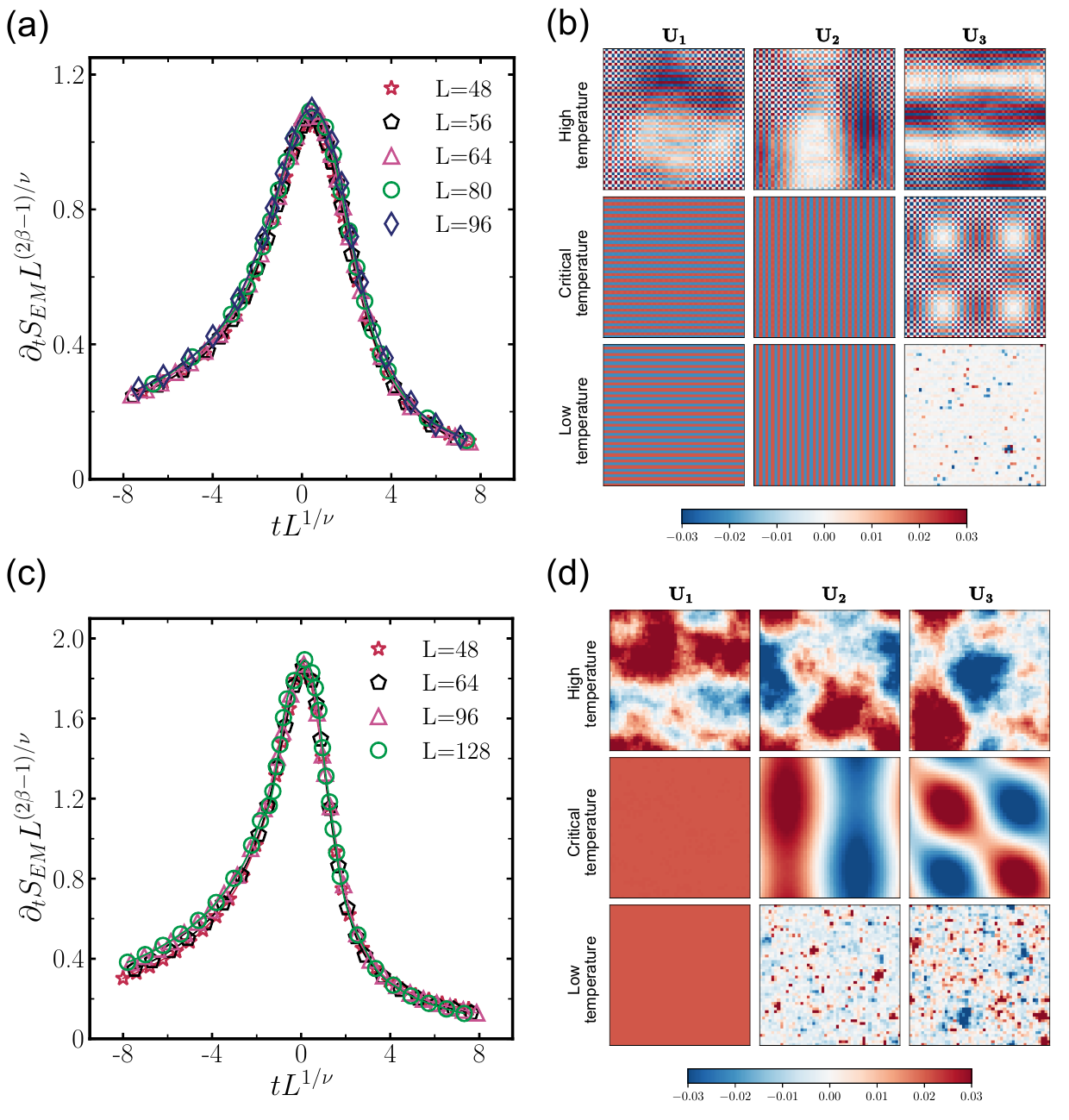}
\centering
\caption{\label{fig:fig2} Eigen microstates and $S_{\text{EM}}$ in frustrated Ising models. (a) Data collapse of $\partial_t{S}_{\text{EM}} L^{(2\beta-1)/\nu}$ versus $tL^{1/\nu}$ with critical exponents $\nu=0.78$, $\beta=0.1$ for the frustrated Ising model with $\kappa = 0.8$. (b) Spatial patterns of the three dominant eigen microstates at high ($T=4$), critical ($T=1.568$), and low ($T=0.8$) temperatures for the frustrated Ising model with $\kappa = 0.8$. (c) Data collapse of $\partial_t{S}_{\text{EM}} L^{(2\beta-1)/\nu}$ versus $tL^{1/\nu}$ with critical exponents $\nu=1$, $\beta=0.125$ for the frustrated Ising model with $\kappa = 0.2$.  (d) Spatial patterns of the three dominant eigen microstates at high ($T=4$), critical ($T=1.608$), and low ($T=0.8$) temperatures for the frustrated Ising model with $\kappa = 0.2$. }
\end{figure}

\subsection{Frustrated Ising model}
We consider the frustrated Ising model to show that $S_{\text{EM}}$ can be used not only for quantitative studies of criticality, but also to identify the dominant microscopic patterns governing the system by analyzing the key eigen microstates associated with significant eigenvalues. The 2D frustrated Ising model with couplings $J_1<0$ (ferromagnetic) and $J_2>0$ (antiferromagnetic) is defined by the Hamiltonian: 
\begin{equation}
 	\mathcal{H} = J_1\sum_{\langle ij \rangle}s_is_j + J_2\sum_{\langle\langle ij \rangle \rangle} s_is_j,   
\end{equation}
where $\langle ij \rangle$ and $\langle\langle ij \rangle \rangle$ denote the nearest and next-nearest (diagonal) neighbors on the square lattice. The ratio $\kappa = J_2 / |J_1|$ quantifies the competition between ferromagnetic and antiferromagnetic interactions, determining the critical behavior of the model: when $\kappa < 1/2$, the system transitions to a ferromagnetic phase; when $\kappa > 1/2$, it transitions to a striped phase (super antiferromagnetic phase). Despite being a simple extension of the standard 2D Ising model, the nature of the stripe transition remains highly controversial, with a central open question being whether the phase transition for $\kappa > 0.5$ is continuous or weakly first-order~\cite{kalz2011analysis,jin2012ashkin,hu2021numerical,li2021tensor}.

We use the standard single-spin Metropolis update combined with a parallel-tempering Monte Carlo algorithm to generate spin configurations (see supplementary data for
details~\cite{liu2024SM}). Taking a typical value of  $\kappa = 0.8$ as an example, the scaling behavior of $\partial_t{S}_{\text{EM}}$ is confirmed in figure~\ref{fig:fig2}(a), supporting that the phase transition is continuous. According to the data fitting of this scaling function, $T_c = 1.5680(4)$,  $\beta = 0.1$ and $ \nu=0.78$ are obtained (shown in the supplementary data~\cite{liu2024SM}), which agree well with the results in Ref.~\cite{kalz2012location,ramazanov2016thermodynamic}. 

The structure of the emerging phase can be extracted from the eigen microstates associated with eigenvalues that contribute significantly to  $S_{\text{EM}}$. Figure~\ref{fig:fig2}(b) shows the three eigen microstates with the largest eigenvalues. At high temperatures, faint horizontal and vertical stripes appear but are smeared by disordered features. Near the critical point, patterns of two significant horizontal and vertical stripe eigen microstates begin to emerge. As the temperature decreases further, these two eigen microstates become dominant. They are the two degenerate ground states, reflecting a super-antiferromagnetic phase driven by an additional next-nearest-neighbor antiferromagnetic interaction, while other states gradually vanish (shown in the supplementary data~\cite{liu2024SM}). The information obtained from the patterns of these eigenstates is a valuable complement to the entropy itself. Otherwise, relying solely on entropy would merely indicate the occurrence of a phase transition, without providing information regarding the nature of the ordered phase. Our $S_{\text{EM}}$, combined with the analysis of its constituent eigen microstates, greatly enhances its interpretability and ability to uncover the mechanisms of phase transitions. This is especially important when the universality class is already determined by the behavior of entropy, but the ordered states require further exploration. When $\kappa=0.2$, the behavior of $S_{\text{EM}}$ is similar to that in $\kappa=0.8$, but the eigen microstates form an Ising-like cluster pattern without stripes (figure~\ref{fig:fig2}(d)). Scaling analysis confirms that the case of $\kappa=0.2$ belongs to the Ising universality class ($\nu=1$, $\beta = 0.125$) (figure~\ref{fig:fig2}(c)).

\subsection{Potts model}
We then study the 2D Potts model ($q=3,4$), described by the Hamiltonian $
H = -J \sum_{\langle ij \rangle} \delta_{\sigma_i, \sigma_j}$,
where $\sigma_i = 1, 2, \dots, q$ and $J>0$ is the nearest-neighbor coupling. As a generalization of the Ising model to more than two states per spin, the Potts models possess $\mathbb{Z}_q$ symmetries, leading to richer critical behaviors and distinct universality classes compared with the Ising case ($q=2$). We conduct FSS analyses on $S_{\text{EM}}$ and $\partial_t S_{\text{EM}}$ (discussed in detail in supplementary data~\cite{liu2024SM}). The calculated critical exponents $(\nu,\beta)$ are consistent with the known exact values, demonstrating that $S_{\text{EM}}$ faithfully captures both the critical behavior and the associated universality class in models with more complex internal symmetries.

\section{Eigen microstate entropy in living systems}

\begin{figure}[!h]
\includegraphics[width=0.8\textwidth]{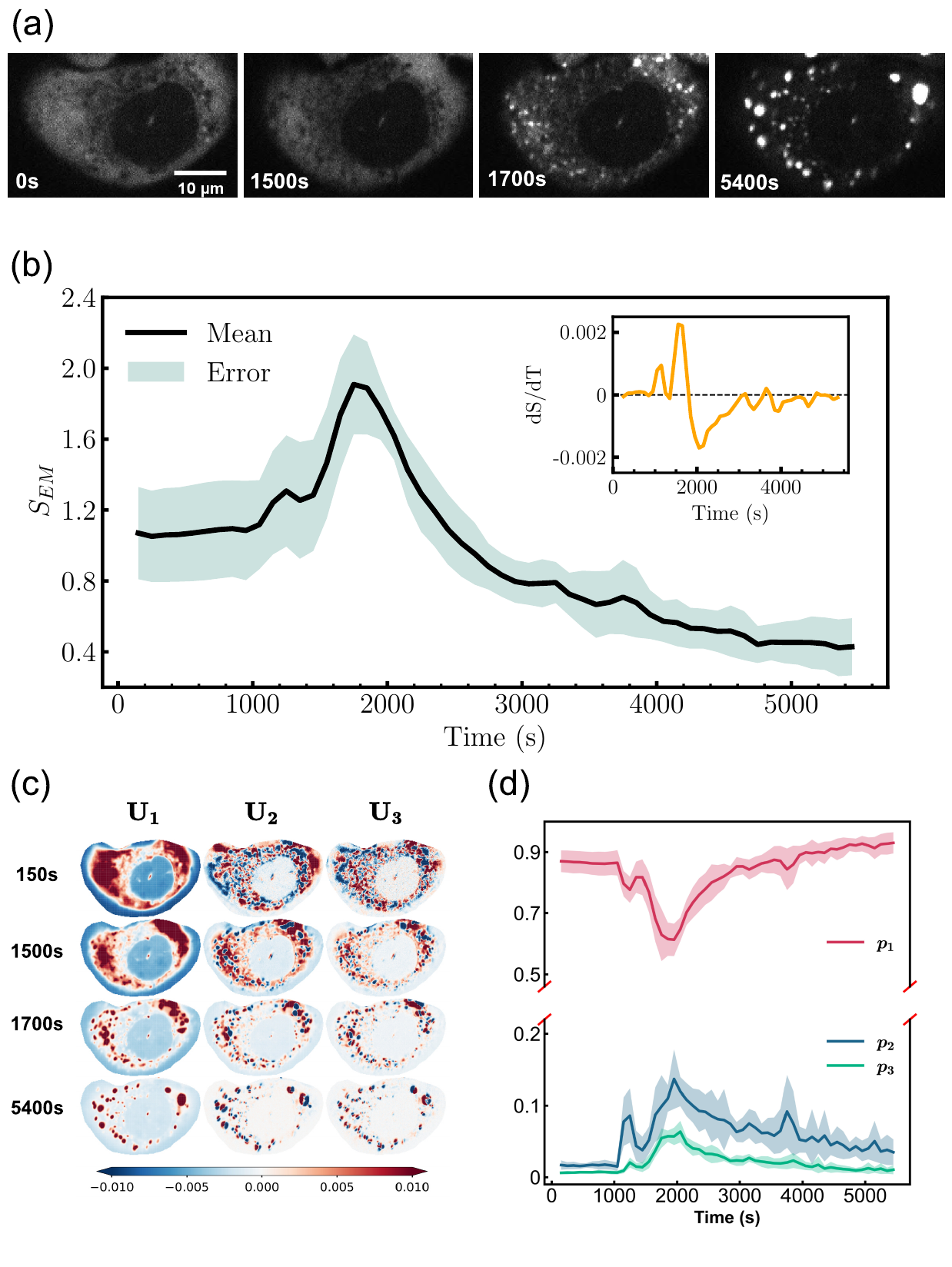}
\centering
\caption{\label{fig:fig3} The phase transition in LLPS is revealed by $S_{\text{EM}}$. (a) Imaging of a U2OS cell with diffusing G3BP1-EGFP. Bright spots correspond to diffusing G3BP1-EGFP molecules imaged at different time points. Camera exposure time is 200 ms. (b) $S_{\text{EM}}$ versus time in living cells (10 samples). The black line gives the mean over all samples, and its shaded region denotes the standard deviation. The inset figure presents the time derivative of entropy versus time. (c) Spatial pattern of the three largest eigen microstates at different times. (d) The probability weights of the three largest eigen microstates versus time.}
\end{figure}

To demonstrate the applicability of $S_{\text{EM}}$ in real-world systems, we apply it to study LLPS in living cells --- a ubiquitous non-equilibrium process where biomolecules assemble into membraneless compartments, playing crucial roles in various cellular functions~\cite{guo2014probing,shin2017liquid,klosin2020phase,demarchi2023enzyme,winter2025phase}. Using stress granules formation in human osteosarcoma cells (U2OS) expressing EGFP-G3BP1 as a LLPS experiment model~\cite{yang2020g3bp1,zhang2024interplay}, we initiate the LLPS by adding sodium arsenite to the cell culture medium, and image the whole process in real time under an inverted spinning-disk confocal microscope, with a total of 5400 s at 1 s intervals (shown in the supplementary data~\cite{liu2024SM}).  The LLPS process is shown in figure~\ref{fig:fig3}(a): Initially, the cells exhibit a homogeneous fluorescence image; small condensates then appear and coalesce into larger droplets after approximately 1700 s. 

Since the fluorescence intensity of each pixel directly reflects the local protein density, we define the microstate as $\bm{s}(\tau)  = [I_1(\tau),\dots,I_N(\tau) ]^{\mathsf{T}}$, where $I_i(\tau)$ denotes the fluorescence intensity of pixel $i$ at time $\tau$, and $N$ is the pixel count. For the time-series data from empirical measurements, such as $\bm{s}(\tau)$ here, we analyze the time evolution of $S_{\text{EM}}$ using a sliding time window. Specifically, we construct an ensemble matrix $\bm{\mathrm{A}}(\tau) = [\bm{s}(\tau-150),\dots, \bm{s}(\tau+149)]$ over a 300-second window. By decomposing the time-dependent ensemble matrix, we track the entropy evolution, as shown in figure~\ref{fig:fig3}(b). Interestingly, unlike the monotonic behavior observed in equilibrium models, biological LLPS exhibits a non-monotonic $S_{\text{EM}}$ curve characterized by an initial increase peaking around 1700 s, followed by a decrease. This behavior is supported by the derivative of $S_{\text{EM}}$ (inset of figure~\ref{fig:fig3}(b)) and is confirmed to be independent of the time-window length (shown in the supplementary data~\cite{liu2024SM}). 

The temporal evolution of $S_{\text{EM}}$ provides insight into the complex process of LLPS. The value of $S_{\text{EM}}$ depends on $\{p_I\}$: a more uniform distribution results in a higher $S_{\text{EM}}$. Initially, the system is dominated by a homogeneous $\bm{U}_1$  (figure~\ref{fig:fig3}(c)) with $p_1 \approx 0.9$, consistent with fluorescent images showing no LLPS droplets (figure~\ref{fig:fig3}(a)). From 0 to 1700 s, $p_1$ decreases while other probabilities like $p_2$ and $p_3$ increase (figure~\ref{fig:fig3}(d)), leading to a more uniform $\{p_I\}$ distribution and resulting in a rising $S_{\text{EM}}$; when $S_{\text{EM}}$ reaches its maximum at 1700 s, the homogeneous $\bm{U}_1$ is replaced by a heterogeneous pattern with small spots (figure~\ref{fig:fig3}(c)), indicating the emergence of a new phase. From 1700 to 5400 s, the gradual rise in new phase ratio, $p_1$, leads to a reduction in $S_{\text{EM}}$. This reduction corresponds to the known biological process where EGFP-G3BP1 condensates grow and aggregate into large droplets (shown in the supplementary data~\cite{liu2024SM}).

The latter stage of biological LLPS and the cooling process of the Ising model both show a decrease in $S_{\text{EM}}$, which results from the growth of the new ordered phase. However, contrary to intuition, the intracellular LLPS exhibits an initial increase in $S_{\text{EM}}$, during which the appearance of small spots in $\bm{U}_1$ (figure~\ref{fig:fig3}(c)) implies that proteins are beginning to spatially correlate without yet forming observable condensates. This finding is consistent with recent experiments where protein clusters were observed prior to condensations~\cite{lan2023quantitative, Kar2022}, which may facilitate protein condensation and play a crucial role in cellular assembly~\cite{pezzotti2023liquid,mukherjee2023thermodynamic}.

\section{$S_{\text{EM}}$ as a precursor signal of El Ni\~no events}
El Ni\~no, in which non-linear interactions play a significant role, is a climate pattern with abnormal warming of the central and eastern equatorial Pacific Ocean that influences the global climate system~\cite{meng2020complexity}. It is typically identified by sea surface temperatures at least $0.5^\circ\text{C}$  above average within the Ni\~no 3.4 region ($5^{\circ}\text{S} - 5^{\circ}\text{N}$, $170^{\circ}\text{W} - 120^{\circ}\text{W}$) for a sustained period~\cite{zhao2024explainable}. We apply $S_{\text{EM}}$ to study the El Ni\~no events using surface air temperature (SAT) data from the ERA5 reanalysis dataset at a spatial resolution of $1^{\circ}\times 1^{\circ}$. We define the microstate as $\bm{s}(\tau) = [T^{a}_1(\tau),\dots,T^{a}_N(\tau)]^{\mathsf{T}}$, where $T^{a}_i(\tau)$ is the anomaly SAT of location $i$ and $N=561$ is the number of grid points (or nodes) within the Ni\~no 3.4 region. An ensemble matrix $\bm{\mathrm{A}}(\tau)$, corresponding to day $\tau$, is constructed using a one-year sliding time window before this day, as: $\bm{\mathrm{A}}(\tau) = [\bm{s}(\tau-364),\bm{s}(\tau-363),\dots, \bm{s}(\tau)]$ (discussed in detail in supplementary data~\cite{liu2024SM}).  By decomposing $\bm{\mathrm{A}}(\tau)$ into eigen microstates $\{ \bm{U}_I \}$ with corresponding probability distribution $\{ p_I \}$, we can obtain the time evolution of $S_{\text{EM}}(\tau)$ according to Eq. \eqref{eq:entropy_define}.

\begin{figure}[!h]
\includegraphics[width=0.9\textwidth]{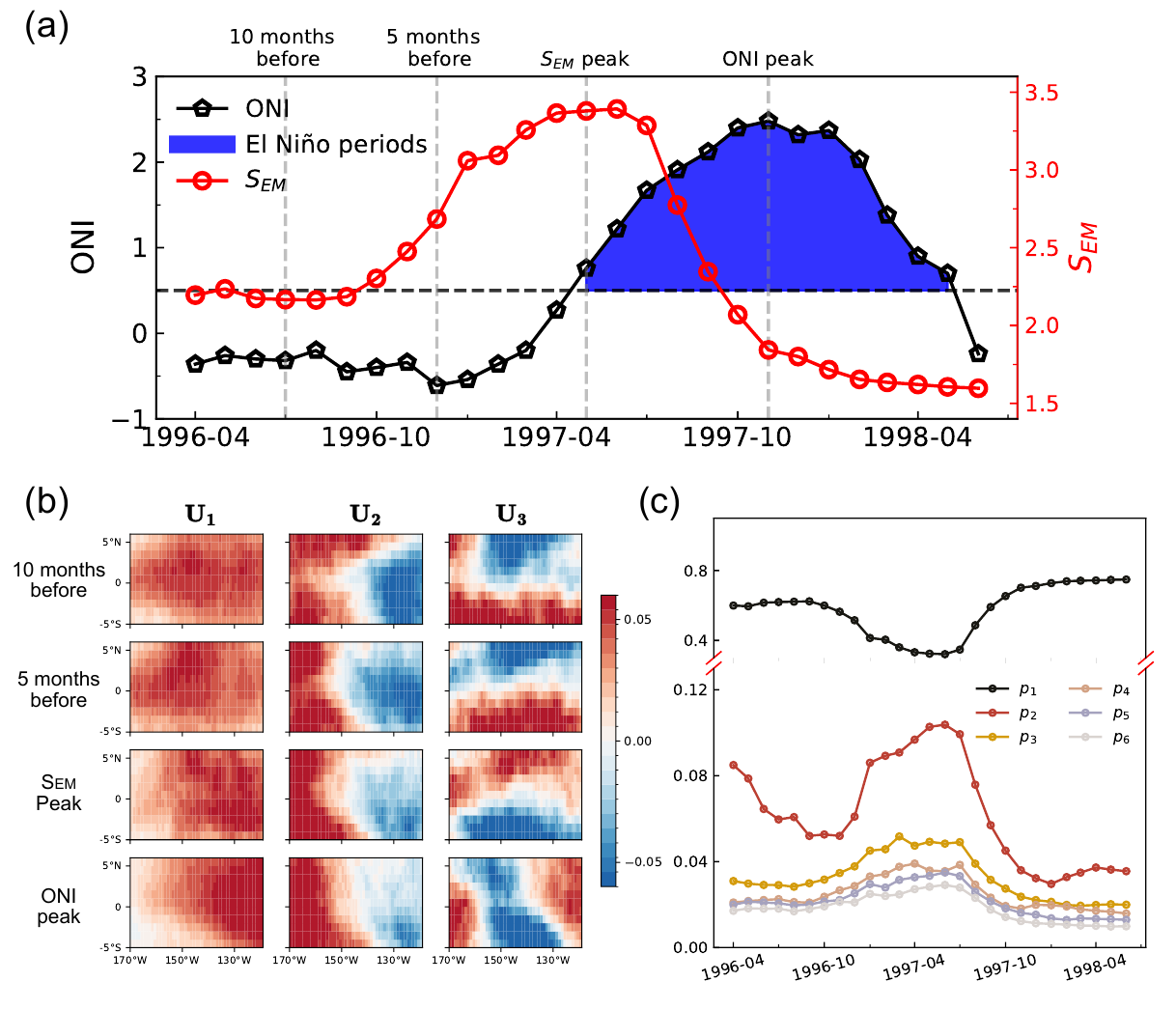}
\centering
\caption{\label{fig:fig4} $S_{\text{EM}}$ serves as a precursor signal for the 1997/1998 El Ni\~no event. (a)  Temporal evolution of $S_{\text{EM}}$ (the red line) alongside the ONI (the black line) of the 1997/1998 El Ni\~no event. Blue shading indicates El Ni\~no periods. (b) Spatial pattern of the three largest eigen microstates at 10 months before (July 1996), 5 months before (December 1996), at the $S_{\text{EM}}$ peak (May 1997), and at the ONI peak (November 1997) of the 1997/1998 El Ni\~no event. (c) The probability weights associated with the six leading eigenvalues as a function of time.}
\end{figure}

We select the extreme 1997/1998  El Ni\~no event, recognized as the strongest Eastern Pacific (EP) El Ni\~no on record~\cite{paek2017were}, as a case study. The Oceanic Ni\~no Index (ONI), defined as the 3-month running mean of anomaly temperature in the Ni\~no 3.4 region, is the primary indicator for identifying El Ni\~no events, with values above $0.5^\circ\text{C}$ typically indicating an El Ni\~no event (see the blue regime in figure~\ref{fig:fig4}(a))~\cite{glantz2020reviewing}. As shown in figure~\ref{fig:fig4}(a), $S_{\text{EM}}$ rises sharply about seven months before the onset of El Ni\~no in May 1997, while it then gradually declines before this  El Ni\~no event ends in May 1998. The increase in entropy that occurs several months before onset can be considered as a precursor signal.

By analyzing the weights of the dominant eigen microstates, we can understand the behavior of entropy increase. In the early phase of El Ni\~no development (approximately 10 months before onset), the system is dominated by a zonally homogeneous $\bm{U}_1$ (figure~\ref{fig:fig4}(b)) with $p_1 \approx 0.6$. As time evolves, $p_1$ decreases significantly, while other probabilities such as $p_2$ and $p_3$ increase (figure~\ref{fig:fig4}(c)), leading to a more uniform distribution of $\{ p_I \}$ and consequently an increasing $S_{\text{EM}}$. When $S_{\text{EM}}$ reaches its peak (in May 1997), the initially homogeneous $\bm{U}_1$ transitions to a zonally heterogeneous pattern (figure~\ref{fig:fig4}(b)). Thereafter, with the gradual increase of $p_1$ and a decline in $S_{\text{EM}}$, $\bm{U}_1$ again becomes dominant, reflecting the system's transition into a new ordered phase as the El Ni\~no event occurs. The emergence of zonal heterogeneity in $\bm{U}_1$ likely reflects localized warming anomalies near the South American coast—a distinctive feature of EP El Ni\~no events~\cite{geng2022emergence}. This structural change may be associated with the destabilization of equatorial waves, particularly downwelling Kelvin waves, which is crucial in the development of El Ni\~no by depressing the thermocline and warming the surface waters in the EP~\cite{mcphaden1999equatorial,wang2018review}. Similar behaviors of $S_{\text{EM}}$ have been observed for other significant El Ni\~no events in recent decades (shown in the supplementary data~\cite{liu2024SM}).

\section{Conclusion}
We have successfully developed an entropy measure, $S_{\text{EM}}$, based on statistically independent eigen microstates. Grounded in the concept of statistical ensembles, this entropy measure plays a foundational role for the investigation of general non-equilibrium statistical systems. Because $S_{\text{EM}}$ is not constrained by the definition of an energy function or specific mathematical formulations, it is highly suitable for studying data-driven complex systems. For model systems, we established well-defined FSS behavior of $S_{\text{EM}}$, verified by exact solutions in the mean spherical model, and applied it to determine the criticality of the frustrated Ising model, a long-standing issue. We have obtained meaningful results in both experimental and empirical systems. In LLPS, the transition from one ordered phase to another involves an intermediate entropy increase. In El Ni\~no events, we observed similar entropy dynamics and demonstrate that $S_{\text{EM}}$ can serve as a precursor signal.

Our entropy definition is based on the concept of eigen microstates and enables the analysis of dominant eigen modes to enhance interpretability beyond entropy signals. It is worth mentioning that the number of eigenvalues of a system is much smaller than the dimension of the phase space, implying the advantage of dimensionality reduction, making it particularly suitable for big-data, high-dimensional complex systems. Moreover, the divergence of relaxation times near a critical point, known as the critical slowing down (CSD), is characterized by a decreasing  $S_{\text{EM}}$, resulting from a more concentrated distribution of $\{p_I\}$ as the system lingers in a smaller subset of states (discussed in detail in supplementary data~\cite{liu2024SM}). Applying our method to study CSD in climate tipping points~\cite{liu2023teleconnections,morr2024detection,boers2025destabilization} and glassy dynamics represents an intriguing direction for future research.

\section*{Acknowledgments}
This work was supported by the National Natural Science Foundation of China (Grant No.~12135003, 12122402, 12475033, 12574221) and the National Key R\&D Program of China
(Grant No.~2023YFE01
09000). G.K.H. acknowledges funding from the China Postdoctoral Science Foundation (Grant No.~2023M730299).

\section*{Author contributions}
T.L., M.X.L, H.L., and X.S.C. conceived and designed the study. X.S.C. derived the analytical results of the mean spherical model. H.L. designed the LLPS experiment. T.L. performed the simulation and computations, and analyzed the results. X.Z.N. and M.L.Z. prepared the samples and performed the LLPS experiment. All authors discussed the
results. T.L., M.X.L, H.L., and X.S.C. wrote the paper with contributions from all authors. X.S.C. supervised the project.

\section*{Supplementary data}
Supplementary material for this article is available online.

\section*{Conflict of interest}
The authors declare no competing interests.

\bibliographystyle{unsrtnat} 
\bibliography{reference.bib}

\end{document}